# Security Architecture for Cluster based Ad Hoc Networks

**Preetida Vinayakray-Jani*,**

DA-IICT, Gandhinanager, India

Email: preeti.vinayakray@gmail.com

*Corresponding Author

**Sugata Sanyal**

School of Technology and Computer Science, Tata Institute of Fundamental Research, Mumbai, India

Email: sanyals@gmail.com

------------------------------------------------------------ABSTRACT------------------------------------------------------------

Abstract- Mobile Ad hoc Networks (MANETs) are subject to various kinds of attacks. Deploying security mechanisms is difficult due to inherent properties of ad hoc networks, such as the high dynamics of their topology, restricted bandwidth, and limited resources in end device. With such dynamicity in connectivity and limited resources it is not possible to deploy centralized security solution but distribution solution. The paper proposes architectural security concept in distributed manner where network is divided into clusters with one cluster head node each. This cluster head node also act as a router providing proactive hidden routing by using Steganographic methods for inter-cluster security. Besides cipher method is used to provide intra-cluster security. The proposed secure architecture specifies operational view of cluster head as a router that provides trust, anonymity and confidentiality through Steganography and Cryptography respectively.

## 1. INTRODUCTION

Mobile Ad hoc Networks (MANETs) allow mobile nodes to form a self-organized network without need for a permanent infrastructure. Such networks can be used in the defence applications, disaster management or in remote locations where supportive fixed network infrastructure is not available. Such MNs are limited by bandwidth, computation and battery life. As a prerequisite to communication, an efficient route between network nodes must be established, and it must adapt to the dynamically changing topology of network. This dynamic property has rendered it vulnerable to various security attacks. Also in MANETs, the privacy issue of MNs become more crucial, as radio resources are shared by all nodes. The node's identity is exposed to the channel eavesdropping. Without trying to secure the identity of the nodes which have an important role in the clustering architecture like security services, the vulnerability can be exploited by attackers to create the Denial of Services (DoS) attacks. Many solutions were proposed in literature to secure MANETs; however, few of them take into account the real MANET's characteristicssuch as: mobility, open network, energy limitation, etc.

This paper proposes security architecture for securing communication in MANET. Proposed approach divides the network into clusters and implements the decentralized certification authority and covert channels to hide routing information. Certification authority solution helps intra-cluster security solution by using cluster-wide symmetric key that is known to all cluster nodes or members. Covert channel solution helps secure communication among Cluster Heads (CHs). Hence proposed secure architecture uses the Steganography and cryptography methods to provide inter-cluster and intra-cluster security.

In this following, brief overview of related work described in section II. Section III provides brief view of Steganography, followed by clustering concept in section IV. Section V provides secure architecture for cluster-based MANET. Finally section VI provides conclusion and future works

## 2. RELATED WORK ON SECURING MANET

Researchers in ad hoc network security have shown various techniques to enhance and fortify the ad hoc routing protocols against various security loopholes and vulnerabilities in the MANET.

Securing ad hoc networks proposed in [1] uses a distributed certification authority based on shared certification key and threshold cryptography. Our approach is based on the same general idea, but it is applied to cluster based network structure.

A specially crafted key sharing algorithm distributing key among all network nodes is proposed in [2, 3], instead of subnets. Signed tickets issued by nodes used for access control. Approach proposed in [4] where every participants issue certificate for other nodes in a web-of-trust manner. Thus each participant stores a number of certificates. The two nodes only communicates if local stores contains a certificate path between them
Securing Clustering Algorithm (SCA) in [5] includes security requirements by using a trust value defining how much any node is trusted by its neighborhood and using the certificate as node's identifier to avoid any possible attack like spoofing. Besides SCA forms a hierarchical structure for both security and routing protocols.

Trusted communication platform for Multi-Agent System (MAS) in [6] shows new proactive hidden routing where agent discovery, route updates and hidden communication are cryptographically independent. Usage of Steganographic techniques enables secure and unrestricted hidden communication among multi agents and is a novel contribution for trusted communication between chosen agents using Steganographic channels. Our approach is based on same general idea but applied to cluster based network architecture.

Due to open medium, dynamic topology, distributed cooperation, in MANET one of attacks called spoofing becomes very common [7]. Hence to prevent spoofing, a network Steganography is used to exploit Internet elements and protocols of the purpose of covertly communicating identity information, armed with simple cryptography [8]. Proposed approach in this paper uses network Steganography for hiding routing information and cryptography to secure intra-cluster security.

## 3. STEGANOGRAPHY

In general, Steganography systems are dedicated to multimedia applications where hidden data is distributed in sound files, images and movies [9, 10, 11, and 12]. Steganographic solutions, located in network protocols are not widely spread, but they rely on usage of optional fields of communication protocols. However this paper t shows the usage of Steganography usage in network protocol

This paper proposes Steganography in distributed router functionality within Cluster Head (CH), provides ability to create covert channels between CHs. This covert channel builds stego-paths between CHs.

## 4. CLUSTER BASED CONCEPT IN MANET

A cluster is one of the ad hoc networks architecture where Cluster Head (CH) responsible of management of cluster and other nodes that are members to particular cluster. Gateways (GWs) manage communication between with adjacent clusters as shown in Figure 1. CHs are responsible to broadcast periodically beacons. These beacons contain administrative information for member nodes of the cluster. Besides GWs broadcast beacons informing their respective clusters about their adjacent clusters.

The inter-clusters communication is ensured by the border nodes. For security reasons, not all border nodes can ensure the link between two clusters but they need to have a high trust level to get the getaway status GW, for more details the reader can refer to the trust model in [5]
System.

## 5. SECURING CLUSTER BASED MANET

In proposed work CHs operate as a Certification Authority (CA) in distributed manner to ensure security for under lying cluster nodes by using cluster-wide symmetric key. A security of communication links within single cluster is provided by symmetric encryption. Through proper key exchange protocol CH and member nodes agree to share common key to prevent eavesdropping. Limiting scope intra-cluster security with symmetric key, this paper is focused to provide secure communication links between clusters of the MANET architecture.

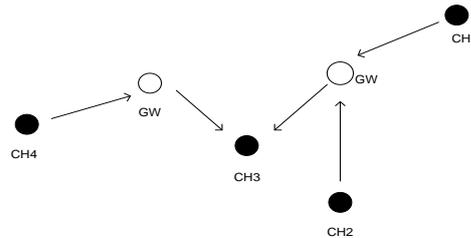

**Figure 1 CH with random walk**

Besides behaving as a CA, CH also performs Steganographic communication in many ways. By using such Steganographic methods on different layers of OSI model. Examples of different layers of OSI model that enable covert channels includes: multimedia and text hiding methods at application layer, transport layer and data link layer.

Cluster gateways act as a distributed Steganographic cluster GW also. Connections are possible between CHs and CG with covert channels. However what Steganographic methods will be used to communicate between CH and CG depend on their stegano-capabilities. With ability to exchange information by using hidden channels all CHs in cluster-based architecture also acts as a distributed router that also supports steganographic methods. Hence CH is acting as a distributed router to carry/covert different covert channels end-to-end. It uses the distance vector protocol without triggered update to prevent potential attacks from random nodes which do not support steganographic methods. It also uses random walk algorithm to perform discovery of new CH that is in vicinity or within communication range. Removed CH with intention of malicious attack is to observe behaviour of other nodes after removal then routing protocol uses triggered updates to inform the change in topology to other nodes. Proposed routing functionality uses distance vector protocol and not link state or hybrid one as they require greater processing time and memory. This routing functionality encounters three operational phases such as: Neighbourhood Discovery phase, Exchanging routing tables and creating secure steg-links and steg- paths.

### 5.1 NEIGHBOURHOOD DISCOVERY

During this operational phase CH sends anonymous message with embedded beacon consists of CH's address and available steganographic methods to use for covert channels. Each CH maintains neighbours and routing table. The former table helps to create steg-link between two CHs that connect them. With created steg-link in neighbourhood table, a periodic hello message is sent to check if the neighbour is still available on steg-link or not. If not then CH refreshes corresponding entry in routing table. If hello message is not received within pre-defined period, the corresponding steg-link gets removed from neighbourhood table.

• Periodic Hello message is triggered to check the existing steg-link. If such Hello message is responded then CH refreshes the entry within the neighbourhood table. If Hello message is not responded within a predefined time interval then such steg-link entry is removed from neighbourhood table entry.

### 5.2 ROUTING UPDATE: EXCHANGING ROUTING UPDATE

CH routing functionality uses covert channels to exchange routing tables between CHs. These updates are sent at regular interval to achieve proactive hidden routing. This proactive routing remains transparent to steganographic connection and discovery phase that is discussed above.

After discovery phase, the new CH's neighbourhood table posses' actual information and it receives entire routing table from neighbouring CH as shown in Figure 4

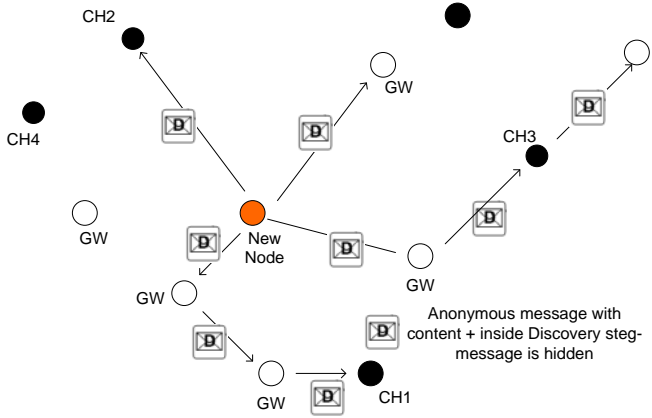

**Figure 2 Discovery phase with random walk algorithm for new CH**

If new CH would like to join this cluster based architecture then during discovery phase following steps will be followed:
• Each CH uses random walk algorithm to discover other CHs in Cluster-based MANET architecture. As shown in Figure 2, new CH attempts to connect the already interconnected CHs
• Each CH and GW passes or drops the discovery steg-message sent based on the random walk algorithm
• Based on the information collected during discovery phase, steg-links are generated between new CH and other CH if their steganographic capabilities are matching
• Two CHs become neighbours if the steg-link exits between them. As a result corresponding steg-link entry is added in new CH neighbourhood table as shown in Figure 3.

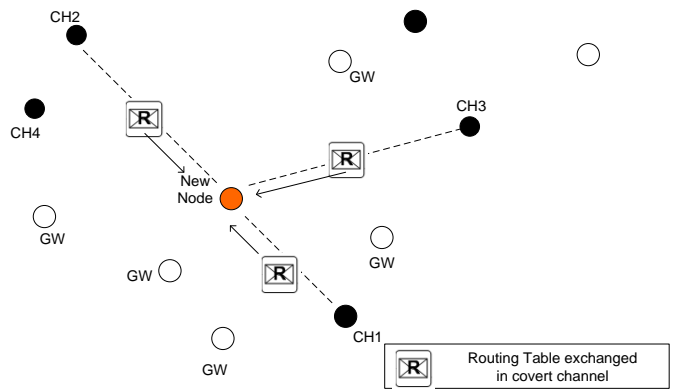

**Figure 4 Exchanging routing information between CHs**

As routing information is exchanged periodically between CHs and when new CH receives the routing tables from its neighbours it is able to learn about other distance CHs and how to reach them. With such information it can also form new steg-links with other CHs.

If one of the CHs terminates or becomes passive or unavailable, the change is directed with hello mechanism. Then routing table updates and change is sent to all the neighbours in the neighbourhood table, when there is time to send the entire routing table.

Each routing entry in the routing table represents the best available steg-path to distance CH with its metric. The metric is based on: 1) Current capacity of the steg-links along end-to-end steg-path; 2) Delay encountered along the steg-path; 3) Common steganographic methods.

The proposed operational phases, expressed in pseudo code are shown in Algorithm 1.

New CH uses random walk algorithm and discovers CH1 and CH2 as shown in Figure 5.

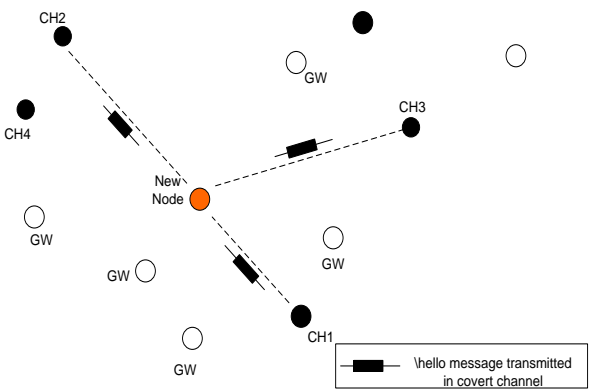

**Figure 3 Creating Steg-links between CHs and updating neighbourhood table of new CH**

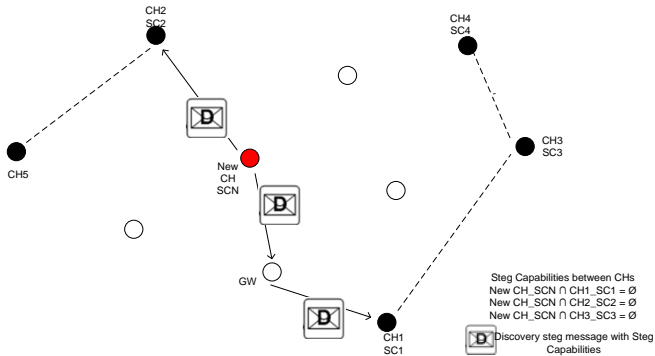

**Figure 5 New CH incompatibility with other CHs**

Nevertheless it is possible that two neighbouring CHs do not have common steganographic methods or capability with new CH. In this situation following functionality can be viewed to improve the convergence as shown in Figure 6.

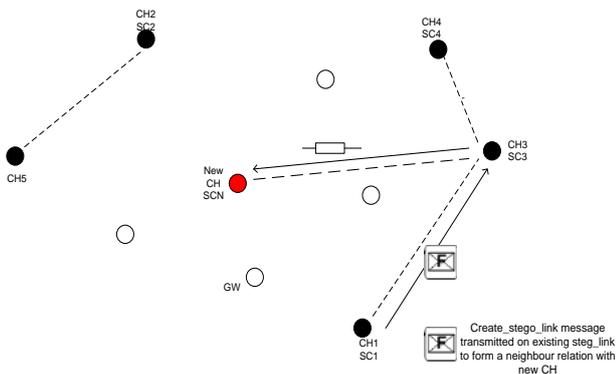

**Figure 6 Mechanism to improve convergence to form steg-links**

In such case the CH will generate Create_steg_link to another CH which possess the steganographic method compatible with new CH. It will send Hello message to new CH and eventually the routing table is exchanged and new CH will learns about the clustering topology through new steg-links and steg-paths. This convergence algorithm is shown Algorithm 2.

## 6. CONCLUSION AND FUTURE WORKS

Paper presents concept of using Steganography in cluster-based architecture that provides ability to create the covert channels between CHs. Paths and links between CHs can be built through steganographic methods to update routing table within CH.

Future work should address the performance analysis of the proposed routing and limitations. These investigations allow an evaluation of proposed security architecture for cluster-based MANET architecture

## Algorithm 1

```
randomWalkRequest ←listenCHs()
routingUpdateRequest ←listenNETWORK()
hello ←listenNETWORK()
do
{
if (randomWalkPeriod + random(fluctuationRW)exceeded)
sendRandomWalk(myAddress, myCovertChannels)
if (routingUpdatePeriod + random(fluctuationRU)exceeded)
sendRoutingUpdate(myRoutingTable)
if (helloPeriod + random(fluctuationH) exceeded)
sendHello(myNeighboursTable)
if (randomWalkRequest)
{
if (findStegMsg(randomWalkRequest))
{
foundAddress, foundCovertChannels ←
uncover(randomWalkRequest)
if (isNewEntry(foundAddress, foundCovertChannels))
{
myRoutingTable ← updateMyRoutes(foundAddress,
foundCovertChannels)
sendRoutingUpdate(myRoutingTable)
}
}
forwardRandomWalk(randomWalkRequest)
}
if (routingUpdateRequest and findChanges(routingUpdateRequest))
{
myRoutingTable ← updateMyRoutes(routingUpdateRequest)
sendRoutingUpdate(myRoutingTable)
}
if (hello)
{
myRoutingTable ← updateNeighborLastHelloTime(hello)
}
for each neighbor ← entry(myNeighborTable)
if(helloTimeout(neighbor)exceeded)
{
myNeighborTable ← removeEntry(neighbor)
sendRoutingUpdate(myRoutingTable)
}
}while (∞)
subroutine sendRandomWalk(address, channels)
{
destination ← selectRandomAgent(myPlatform)
sendViaMAS(destination, cover(address,channels))
}
subroutine forwardRandomWalk(message)
{
if (coinFlip(pf) = heads)
{
destination ← selectRandomAgent(myPlatform)
sendViaMAS(destination, message)
}
}
subroutine sendRoutingUpdate(table)
for each destination ← entry(myNeighborTable)
sendViaNETWORK(destination, cover(table))
}
```

**Algorithm 2**
```
if (newDataToSend)
{
paths ← findPathsMatch(myRoutingTable,
destination)
if (count(paths) > 1)
{
calcMetricsForPaths(paths, capacity, delay,
steg_method)
BPath ← chooseBestPath(paths)
sendData(BPath)
}
```

```
else
if (count(paths) = 1) sendData(paths)
else noPathFound()
}
```